\documentstyle[12pt]{article}
\begin{document}
\centerline{\large{\bf Interpreting the time of decay 
measurement: phenomenological}}
\vskip 0.2in
\centerline{\large{\bf significance of the Bohm model}}

\vskip 0.5in

\centerline{A. S. Majumdar\footnote{e-mail:archan@boson.bose.res.in}}
\vskip 0.2in
\centerline{S.N.Bose National Centre for Basic Sciences}
\vskip 0.1in
\centerline{Block-JD, Sector III, Salt Lake, Calcutta 700098,
India.}
\vskip 0.2in
\centerline{Dipankar Home\footnote{e-mail:dhom@boseinst.ernet.in}}
\vskip 0.2in
\centerline{Bose Institute, Calcutta 700009, India}

\vskip 2.0in

\begin{abstract}

We point out that the spreading of wave packets could be
significant in affecting the analysis of  experiments involving the
measurement of time of decay. In particular, we discuss a hitherto
unexplored application of the Bohm model in properly taking into account
the nontrivial effect of wave packet spreading  in the CP violation
experiment.

\end{abstract}

PACS No. 03.65.Bz

\pagebreak

In quantum mechanics, if a particle is represented by a wave packet
which evolves according to the Schrodinger equation, an
inevitable consequence is the spreading of such a  packet. In
most experiments the effect of wave packet spreading is
negligible in considering the outcome of the measurement.
However, there may exist situations where the magnitude
of wave packet spreading, in spite of being  small, could be
nonnegligible when compared to the magnitude of  delicate
quantum effects being studied. In particular, the observations of
minute quantum effects concerning the decay of unstable particles
which are inferred from observing the times of decay could be
affected by wave packet spreading. This is because the time of
decay cannot be measured directly and has to be inferred from
appropriate measurements on the decay products whose packets spread
before reaching the relevant detectors. Such situations, for
example, could occur in verifying CP violation~[1], or verifying the
departure from exponential decay of unstable nuclei for very short
times~[2]. In this paper we focus on the CP violation experiment and
analyse closely as to what extent the wave packet spreading affects
the inference of CP violation from the performed measurements. We argue
that, contrary to the usual belief, one  {\it does} require to take into
account the  wave packet spreading in order to infer CP violation
precisely. Our analysis also reveals an interesting application of 
the Bohm model
in quantifying the effect of wave packet spreading in such
experiments.
 
At the outset, let us note that the time of decay is not
represented by any hermitian operator. However,
in quantum mechanics, any standard measurement is described in terms
 of hermitian operators. Even for an observable for
which there exist no directly corresponding hermitian operator,  measurement
of it can  ulimately be reduced to the measurement of some hermitian
operator. For example, when one measures  the wavelength 
of light, one is actually measuring 
position  corresponding to  fringe
spacing. Other quantities like mass are
also finally measured in terms of positions noted from
mass spectrograph. 
The usual textbook justification as to why a nonhermitian operator is not
measurable is that its eigenvalues are, in general, complex. However, this is
not an adequate argument because the real and/or imaginary
components can be taken to
correspond to observables. This occurs, for instance, in decay processes 
where the effective Hamiltonian is complex with its real and imaginary
parts corresponding to mass and decay rate respectively. An example of
particular relevance is the physics
of the nonorthogonal $|K_L>$ (long-lived kaon) and $|K_S>$ (short-lived
kaon) states which are eigenstates of a non-hermitian effective
Hamiltonian. The experimentally measured distinction between these
two states is crucial for the inference of CP-violation in weak interactions
of particle physics. On the other hand, there is a theorem based on 
general quantum mechanical considerations that any measurement of
a nonhermitian operator with nonorthogonal eigenstates would lead to
superluminal signalling using the EPR-Bohm type  nonlocal correlations~[3].

Here we recall a simple general argument ruling out the measurability
of a nonhermitian operator if a measurement process is describable by
linear and unitary quantum mechanics.
 Let $|\psi_1>$ and $|\psi_2>$ be the
eigenstates of an observable. The measured state is, say,
\begin{eqnarray}
|\Psi> = a|\psi_1> + b|\psi_2>
\end{eqnarray}
 and the apparatus is in an initial state
$|A_0>$. Linearity demands that the  system-apparatus state $|\Psi>|A_0>$ 
 evolves by measurement interaction as
\begin{eqnarray}
a|\psi_1>|A_0> \rightarrow  a|\psi_1>|A_1> \nonumber \\
b|\psi_2>|A_0> \rightarrow  b|\psi_2>|A_2>
\end{eqnarray}
From unitarity, it then follows 
\begin{eqnarray}
a^{*}b<\psi_1|\psi_2> = a^{*}b<\psi_1|\psi_2><A_1|A_2>
\end{eqnarray}
Now, if the states $\psi_1>$ and $\psi_2>$ are
nonorthogonal eigenstates of a nonhermitian operator, i.e.,
$<\psi_1|\psi_2> \neq 0$, then Eq(3) cannot be satisfied for any measurement
in which the two  distinguishable states $|A_1>$ and $|A_2>$ 
are orthogonal. Hence, according to standard quantum mechanics,
nonhermitian operators cannot be measured if  measurement
processes are linear and unitary.

Now, turning to the particular case  of measuring the
time of decay, it has to be
inferred from the measurement of an appropriate hermitian observable. Usually
this involves measurement of position/momentum of decay products. In the
theory   of scattering and decay
processes, one  describes the decaying particles as well as the decay
products in the asymptotic limit by plane waves.
However, to be more realistic, one needs to use
 wave packets instead of plane waves. We shall now argue that
 the spreading of wave packets in such an experiment involving CP
violation gives rise to considerable
departures from the predicted  coordinates of the decay products.

First, let us recapitulate a few basic features of CP-violation~[4].
C(charge conjugation) and P(parity) are two of the fundamental
discrete symmetries of nature, the violations of which have not
been empirically detected in phenomena other than weak interactions.
If a third discrete symmetry T(time reversal) is
taken into account, there exists a fundamental theorem of quantum
field theory, viz., the CPT theorem which states that all physical
processes are invariant under the combined operation of CPT.
However, there is no theorem forbidding the violation of CP
symmetry. In fact, there have been several experiments to
date~[5], starting from the pioneering observation of Christenson,
Cronin, Fitch and Turlay~[4], that have revealed the occurrence of CP
violation through weak interactions  involving
the particles $K^0$ and $\bar{K^0}$. The eigenstates of strangeness
$K^0$ \hskip 0.1in $(s=+1)$ and its CP conjugate $\bar{K^0}$ \hskip
0.1in $(s=-1)$ are produced in strong interactions, for example,
the decay of $\Phi$ particles. Weak interactions do not conserve
strangeness, whereby $K^0$ and $\bar{K^0}$ can mix through
intermediate states like $2\pi, 3\pi, \pi\mu\nu, \pi e\nu$, etc. The
observable particles, which are the long lived $K$-meson $K_L$, and
the short lived one $K_S$, are linear superpositions of $K^0$ and
$\bar{K^0}$, i.e.,
\begin{eqnarray}
\vert K_L \rangle & = & (p\vert K^0\rangle - q\vert
\bar{K^0}\rangle )/ \sqrt{\vert p\vert^2 + \vert q\vert^2} \\
\vert K_S \rangle & = & (p\vert K^0\rangle + q\vert
\bar{K^0}\rangle )/ \sqrt{\vert p\vert^2 + \vert q\vert^2}
\end{eqnarray}
which obey the exponential decay law $\vert K_L\rangle \rightarrow
\vert K_L\rangle exp(-\Gamma_L t/2)exp(-im_Lt)$ and analogously for
$\vert K_S\rangle$, where $\Gamma_L$ and $m_L$ are the decay width
and mass respectively of the $K_L$ particle. It follows from (4)
and (5) that
\begin{eqnarray}
\langle K_L\vert K_S\rangle =  {|p|^2-|q|^2 \over |p|^2+|q|^2}
\end{eqnarray}

CP violation takes place if the states $\vert K_L\rangle$ and
$\vert K_S\rangle$ are not orthogonal. Through weak interactions
 $K_S$  decays rapidly into channels such as $K_S
\rightarrow \pi^{+}\pi^{-}$ and $K_S \rightarrow 2\pi^{0}$  with a
mean lifetime of $10^{-10}s$, whereas the predominant decay modes
of $K_L$ are $K_L \rightarrow \pi^{\pm}e^{\pm}\nu$ (with branching
ratio $\sim 39\%$), $K_L \rightarrow \pi^{\pm}\mu^{\pm}\nu (\sim
27\%)$, and $K_L \rightarrow 3\pi ( \sim 33\%)$ [5]. The CP
violating decay mode $K_L \rightarrow 2\pi$ is extremely rare (with
branching ratio $\sim 10^{-3}$) in the background of the other
large decay modes.  The momenta and
locations of the emitted pions are important since
the key experimental signature is to detect the $2\pi$
particles coming from the decay of $K_L$ and identify them
as coming from $K_L$ and {\it not} from $K_S$.

In a typical experiment to detect CP violation, an initial state of
the type
\begin{eqnarray}
\vert\psi_i\rangle = (a\vert K_L\rangle + b\vert K_S\rangle )
\end{eqnarray}
is used which is a coherent superposition of the $K_L$ and $K_S$
states. Such a state is produced by the technique of
`regeneration'~[6] which is used in a large number of
experiments~[5]. The common feature of all these experiments is the
measurement of the vector momenta $\stackrel{\rightarrow}{p_i}$ of
the charged decay products $\pi^{+}\pi^{-}$ or $2\pi^0$ from the
decaying pions. It is only the {\it type} of instrument used for actually
measuring the momenta that varies from experiment to experiment.

We consider a single event in which the two emitted pions from a
decaying kaon are detected by two detectors respectively along two
different directions. From the measured momenta
$\stackrel{\rightarrow}{p_1}$ and $\stackrel{\rightarrow}{p_2}$,
the {\it trajectories} followed by the individual pions are
{\it retrodictively} inferred {\it assuming} that they have followed 
{\it classical 
 trajectories}.  The point of intersection of these
retrodicted {\it trajectories} is inferred to be the point from which
the decay products have originated from the decaying system. In other
words, what is technically known as the ``decay vertex'' is
determined in this way.  Then momentum of the decaying
kaon is obtained by $\stackrel{\rightarrow}{p_k} =
\stackrel{\rightarrow}{p_1} + \stackrel{\rightarrow}{p_2}$. Once
the decay vertex and the kaon momentum are known, one estimates the
time taken by the decaying kaon to reach the decay vertex from the source,
{\it again} using at this stage the idea of a {\it classical  
trajectory}. If this time turns out to be significantly larger than the
$K_S$ mean lifetime ($\sim 10^{-10}s$), one infers that the
detected $2\pi$ pair must have come from $K_L$ and {\it not} from
$K_S$ which, as already
mentioned, is the signature of CP violation.

It is thus evident from the above discussion that the assumption
of a {\it classical trajectory} of a freely evolving particle
(decaying kaon as well as pion) is a key ingredient in inferring
CP violation in such
 experiments (see also an earlier
paper~[7], but there the  point about the wave packet
description and its spreading was not considered).
Now, to consider the justification of such an assumption we note
that
 within the standard interpretation of quantum
mechanics, the very concept of a trajectory of a particle is
regarded to be inadmissible.
  One possible argument could be to associate localized wave
packets with the emitted pions and decaying kaons, 
and to use the fact that their
peaks follow {\it classical trajectories} in the case of  free
evolutions. 
But then there would
be inevitable spreading of these wave packets.  It is
thus important to consider
 a quantitative estimate of this behaviour for  the experiment
discussed here.
Let us {\it quantify} the
resulting error or fluctuation due to the spreading of a wave packet
by taking into account the {\it
actual distances} involved in the relevant experiments. 
This is particularly crucial in the present context
because the CP violation effect is exceedingly {\it
small}  (branching ratio of the CP violating decay mode $K_L
\rightarrow 2\pi$ is $10^{-3}$). 
We take dimension of the wave packet of the kaon at the time
of its production to be typically of  the order of $1 f$, i.e.,
its initial spread $\sigma_0 \approx 10^{-13}cm$. After a time $t$,
its spread $\sigma$ is given by
\begin{eqnarray}
\sigma = \sigma_0\biggl(1 + ({\hbar t \over 2m\sigma_0^2})^2\biggr)^{1/2}
\end{eqnarray}
After  the kaon  travels about $300 K_S$
decay lengths $(\approx 10 m)$, which is the usual distance involved
in the relevant experiments,  one obtains $\sigma = 10^{5}m$ -- an astonishingly
large number ! One may attribute this to the
nonrelativistic nature of the above analysis. However, 
the prediction of CP violation in
the relevant experiments can be formally described adequately in terms of 
the Schroedinger equation
(see [1] and references therein). Nevertheless, even if
relativistic corrections alter
the value of $\sigma$ by several orders of magnitude, $\sigma$ would
 still be too large for unambiguously inferring CP violation
in such experiments.
Surprisingly, in the analysis of {\it none} of the CP violation
experiments performed to date,  this point has been considered. 

In order to  take into account such an  effect of wave packet
spreading,  the Bohm model (BM) can play a useful role in 
  estimating accurately
the position of decay vertex from the observed position/momentum of
decay products. To explain this, we first briefly recapitulate the key
elements of BM. BM provides an  ontological and
self-consistent interpretation of the formalism of
quantum mechanics in terms of {\it particle trajectories} ~[8,9,10]. 
Predictions of BM 
 are in agreement with that of standard quantum mechanics. In  BM 
a wave function $\psi$ is  taken
to be an incomplete specification of the state of an individual
particle. An objectively real ``position'' coordinate 
(``position'' existing irrespective of any external
observation) is
ascribed to a  particle apart from the wave function. Its  ``position''  
evolves with time obeying an equation that can be
derived from the Schroedinger equation (considering
the one dimensional case)
\begin{eqnarray}
i\hbar {\partial\psi \over \partial t} = H\psi \equiv - {\hbar^2
\over 2m} {\partial^2 \psi \over \partial x^2} + V(x)\psi
\end{eqnarray}
by writing
\begin{eqnarray}
\psi = Re^{iS/\hbar}
\end{eqnarray}
and using the continuity equation
\begin{eqnarray}
{\partial \over \partial x} (\rho v) + {\partial\rho \over \partial
t} = 0
\end{eqnarray}
for the probability distribution $\rho(x,t)$ given by
\begin{eqnarray}
\rho = \vert \psi \vert^2.
\end{eqnarray}
It is important to note that $\rho$ is ascribed an {\it
ontological}
significance by regarding it as representing the probability
density of ``particles'' occupying {\it actual} positions. On the other
hand, in the standard interpretation, $\rho$ is interpreted as the
probability density of {\it finding} particles around certain
positions. Setting ($\rho v$) equal to the quantum probability current
leads  to the Bohmian equation of motion where the particle
velocity $v(x,t)$ is given by
\begin{eqnarray}
v \equiv {dx \over dt} = {1\over m}{\partial S \over \partial x}
\end{eqnarray}
The particle trajectory
is thus deterministic  and is obtained by integrating (13)
for a given initial position.

Now let us examine  the nature of Bohmian trajectories
in the case of a wave packet. An ensemble of particles distributed
over a wave packet
possess different ontological positions. It can be shown
that  particles initially at the centre of the wave packet follow
classical trajectories~[9]. 
But all particles with initial positions 
away from the centre of the wave packet follow {\it nonclassical}
Bohmian trajectories. 
The position of any  particle at a
time $t_2$, denoted by $X(t_2)$, 
can be  computed given its initial position $X_0$
and velocity $v_0$ at time $t_1$. The magnitude of departure 
from classical trajectories
 is embodied in the second term of the following relation (see Ref.~[9])
\begin{eqnarray}
X(t_2) = v_0(t_2 - t_1) + X_0\biggl(1 + ({\hbar (t_2 - t_1) 
\over 2m\sigma_0^2})^2\biggr)^{1/2}
\end{eqnarray}
In any experiment involving the measurement of time of decay, in
particular, in the CP violation experiment, one could
measure  the set of times at which the vector momenta of the
decay products are recorded by a set of detectors located at
various positions. In terms of Eq.(14), the values of $X(t_2)$ are
recorded for various $t_2$. Then inverting Eq.(14), it is
possible to express $X_0$ as a function of $t_1$ for any particular
$t_2$. Using the whole set of such data this exercise can be
used to obtain an ensemble  of retrodicted trajectories for the
decay products. The
various intersection points of these retrodicted trajectories 
correspond to  various decay vertices of the corresponding
subensembles of trajectories. In this way one can estimate a spread of
decay vertices for the kaons. From this, it is possible to obtain
the spread in decay times for the kaons by using the trajectory
equation for kaons, knowing their origins.
 
We have thus shown that using  BM  one can {\it retrodict} the 
trajectories for the decay products and decaying kaons, given the measured vector
momenta and arrival times of the decay products. This would then
enable to calculate the spread in decay times for the decaying
kaons.
The {\it nonclassical} second term in the trajectory equation(14) is crucial
in calculating this spread.
 On the other hand, within the standard framework, the best one can
do is to describe the propagation of the peak of a wave packet by
the classical equation.  In the absence of
any equation of motion for trajectories within standard quantum mechanics,
it then remains a nontrivial issue to take into account consistently the effect
of wave packet
spreading for
 estimating accurately {\it when} the kaons have decayed
into $2\pi$ pairs that  originate only from $K_L$ and not from
$K_S$. BM turns out to provide a convenient scheme for addressing
this issue. It should be
worthwhile to look for more such  examples where BM can usefully 
supplement the standard framework for analysing the
experimental results with more precision.
  An interesting area for such a study would be to 
analyse in terms of Bohmian trajectories 
the recently claimed  experimental detection of minute departure from
the exponential behaviour in quantum tunelling~[11] which  involves
essentially the measurement of decay times.

\vskip 0.5in

The research of DH is supported by the Department of Science and
Technology, Govt. of India.

\pagebreak

{\bf REFERENCES}

\vskip 0.5in

\begin{description}

\item[[1]] For a review, see for instance, K.Kleinknecht, in ``CP
violation'', edited by C.Jarlskog, (World Scientific, Singapore,
1989) pp. 41 -104.
\item[[2]] T. D. Nghiep, V. T. Hanh and N. N. Son,
Nucl. Phys. B (Proc. Suppl.) {\bf 66} (1998) 533.
\item[[3]] E. J. Squires, Phys. Lett. A {\bf 130}, 192 (1988).
\item[[4]]  J.H.Christenson, J.W.Cronin, V.L.Fitch and R.Turlay, 
Phys. Rev. Lett. {\bf 13}, 138 (1964).
\item[[5]] For eample, see, C.Geweniger et al., Phys. Lett. B {\bf
48} (1974) 487; V.Chaloupka et al., Phys. Lett. B {\bf 50} (1974)
1; W.C.Carithers et al., Phys. Rev. Lett. {\bf 34} (1975) 1244;
N.Grossmann, et al., Phys. Rev. Lett., {\bf 59} (1987) 18.
\item[[6]] A.Pais and O.Piccioni, Phys. Rev. {\bf 100} (1955) 1487.
\item[[7]] D. Home and A. S. Majumdar, Found. Phys. {\bf 29}, 721 (1999).
\item[[8]] D.Bohm, Phys. Rev. {\bf 85} (1952) 166; D.Bohm and
B.J.Hiley, ``The Undivided Universe'', (Routledge, London, 1993).
\item[[9]] P.R.Holland, ``The Quantum Theory of Motion'', (Cambridge
University Press, London, 1993).
\item[[10]] 
J.T.Cushing, ``Quantum Mechanics -
Historical Contingency and the Copenhagen Hegemony'', (University
of Chicago Press, Chicago, 1994); D. Home, ``Conceptual Foundations of
Quantum Physics'' (Plenum, NY, 1997).
\item[[11]] S. R. Wilkinson et al., Nature {\bf 387} (1997) 575.

\end{description}

\end{document}